\documentclass[9pt,a4paper,twocolumn]{article}
\usepackage{mathptmx,graphicx}
\usepackage[utf8]{inputenc}
\usepackage{dcolumn}
\usepackage{bm}
\usepackage{braket}
\usepackage[T1]{fontenc} 
\usepackage{float}
\usepackage{amsmath,amsfonts,amssymb,eqnarray}  
\usepackage{color, soul}
\usepackage{graphicx}  
\usepackage{dblfloatfix} 
\usepackage{array,booktabs}
\usepackage[pdftex,hypertexnames=false]{hyperref} 
\usepackage[affil-it]{authblk}
\usepackage[font={small}]{caption}
\usepackage{cite}
\usepackage[left=2cm,right=2cm,top=1.6cm]{geometry}
\usepackage{titlesec}
\titleformat{\section}
  {\normalfont\fontsize{13}{15}\bfseries}{\thesection}{0.5em}{}
\titleformat{\subsection}
  {\normalfont\fontsize{11}{15}\bfseries}{\thesubsection}{0.5em}{}
\titleformat{\subsubsection}
  {\normalfont\fontsize{10}{15}\bfseries}{\thesubsubsection}{0.5em}{}  
\usepackage{authblk}

\title{Quantum optics with single molecules in a three-dimensional polymeric platform}
\author{Maja Colautti$^{1,2}$}
\author{Pietro Lombardi$^{3}$}
\author{Marco Trapuzzano$^{2}$}
\author{Francesco S. Piccioli$^{3}$}
\author{Sofia Pazzagli$^{2,3}$}
\author{\newline Bruno Tiribilli$^5$}
\author{Sara Nocentini$^{1,3}$}
\author{Francesco S. Cataliotti$^{1,2,3}$}
\author{Diederik Wiersma$^{1,2,6}$}
\author{\newline Costanza Toninelli$^{1,3}$*}
\affil{$^1$\small European Laboratory for Non-Linear Spectroscopy (LENS), Via Nello Carrara 1, Sesto F.no 50019, Italy}
\affil{$^2$ \small Dipartimento di Fisica e Astronomia, Università degli Studi di Firenze, Via G. Sansone 1, Sesto Fiorentino 50019, Italy}
\affil{$^3$ \small National Institute of Optics (CNR-INO), Largo Enrico Fermi 6, Firenze 50125, Italy}
\affil{$^5$ \small Institute for Complex Systems (CNR-ISC), Via Madonna del Piano 10, Sesto F.no 50019, Firenze, Italy}
\affil{$^6$ \small Istituto Nazionale di Ricerca Metrologica (INRiM), IT-10135 Torino, Italy}
\affil{$^*$\small toninelli@lens.unifi.it}

\date{\today}

\begin{document}

\twocolumn[ 
\begin{@twocolumnfalse}
\maketitle
     \vspace{-0.8cm}
  \begin{abstract}
      \normalsize
         \vspace{1pt}

The successful development of future photonic quantum technologies heavily depends on the possibility of realizing robust, reliable and, crucially, scalable nanophotonic devices. In integrated networks, quantum emitters can be deployed as single-photon sources or non-linear optical elements, provided their transition linewidth is broadened only by spontaneous emission. However, conventional fabrication approaches are hardly scalable, typically detrimental for the emitter coherence properties and bear limitations in terms of geometries and materials. Here we introduce an alternative platform, based on molecules embedded in polymeric photonic structures. Three-dimensional patterns are achieved via direct laser writing around selected molecular emitters, which preserve near-Fourier-limited fluorescence. By using an integrated polymeric design, record-high photon fluxes from a single cold molecule are reported. The proposed technology allows to conceive a novel class of quantum devices, including integrated multi-photon interferometers, arrays of indistinguishable single photon sources and hybrid electro-optical nanophotonic devices.

\end{abstract}
\vspace{.5cm}  
  \end{@twocolumnfalse}]

Polymers are used in a wide application range, from nanotechnology and biomedicine, to clothing and food packaging\cite{Li2002,Green2016}, being low-cost mass-produced materials with tailored physical, electronic and optical properties. They provide expanded functionalities to optoelectronic and photonic devices~\cite{tang, PAQUET}. 
In optical integrated circuits e.g., polymer photonics extends the design options to three-dimensions (3D) and free-standing structures, supporting hybrid integration~\cite{Schumann2014} 
and interchip connection~\cite{Lindenmann:12}. On the other hand, future technologies rely on light manipulation at the single photon level, targeting unmatched functionalities based on quantum effects. 
This sets new requirements to traditional photonic platforms, such as the integration of quantum emitters, which hold promise as on-demand sources of non-classical light~\cite{Lounis2005} 
and for measurements beyond the classical limit~\cite{Degen2017}. An effective coupling to quantum emitters is indeed mandatory to guarantee efficient collection and processing of the quantum states~\cite{Benson2011}. However, structuring the environment around quantum emitters in the wavelength scale affects the emission properties, typically degrading its quantum coherence and photostability~\cite{Jantzen2016,Liu2018}, thus limiting on-chip operation. 
Furthermore the nanofabrication of such semiconductor hybrid devices \cite{Sipahigil2016, Thyrrestrup2018} is demanding and hardly scalable, while the performances result very sensitive to imperfections. Seminal results reported in Refs.\cite{Shi2016, morozov2018,Sartison2017} point at the key advantages of single-emitter integration in polymeric microstructures, such as one-step 3D patterning and a-priori positioning of the emitter in high-intensity regions of the electro-magnetic field. Nevertheless, the colloidal systems employed in Refs.\cite{Shi2016, morozov2018} are not suitable for most quantum applications due to decoherence and spectral diffusion \cite{Jantzen2016, Frantsuzov2008}, whereas the epitaxially grown quantum dots in \cite{Sartison2017} are not favourable for a full 3D integration. 

Single organic molecules of polyaromatic hydrocarbons (PAHs) have shown remarkable quantum optical properties within numerous combinations of guest chromophores and host matrices, and are potentially compatible with polymer chemistry. In particular, the realization of Anthracene nanocrystals doped with Dibenzoterrylene molecules (DBT:Ac NCXs) exhibiting photostable single-photon emission and Fourier-limited line-widths at $3\,$K \cite{Pazzagli2018}, has opened the pathway to their full embedding in polymeric devices \cite{Schadler2019, Ciancico2019}.

In this work we introduce a hybrid polymeric photonic platform, that shows emission close to the life-time limit from single molecules coupled to 3D integrated optical elements. The nanophotonic chips are obtained by direct laser writing (DLW) of polymers \cite{Kawata2001}, with unique advantages in terms of versatility, large offer of multi-purpose photoresists, sub-micron resolution~\cite{ZHANG2010} and computer-aided manufacturing. 
The compatibility with molecules and broad flexibility in the design
is demonstrated here by embedding Anthracene nanocrystals doped with Dibenzoterrylene molecules at variable heights and well-defined locations in different architectures, fabricated either on silica or on gold substrates. Critical aspects like the performances of the hybrid photonic device under cryogenic cooling and the effect of the whole fabrication process on the emitter quantum coherence are investigated. In particular, close-to Fourier limited emission is observed from on-chip molecules at cryogenic temperatures and enhanced light extraction is achieved in a micro-dome solid immersion lens design. 
These results show how an all-organic platform can be realised, that offers unique solutions for photonic quantum technologies, combining on a chip the freedom of three-dimensional polymeric architectures with the optimal properties of single photon emission from fluorescent molecules. 

\section*{Organic quantum emitters in 3D polymeric structures}
\begin{figure*}[ht!]
\centering
\includegraphics[width=0.85\textwidth]{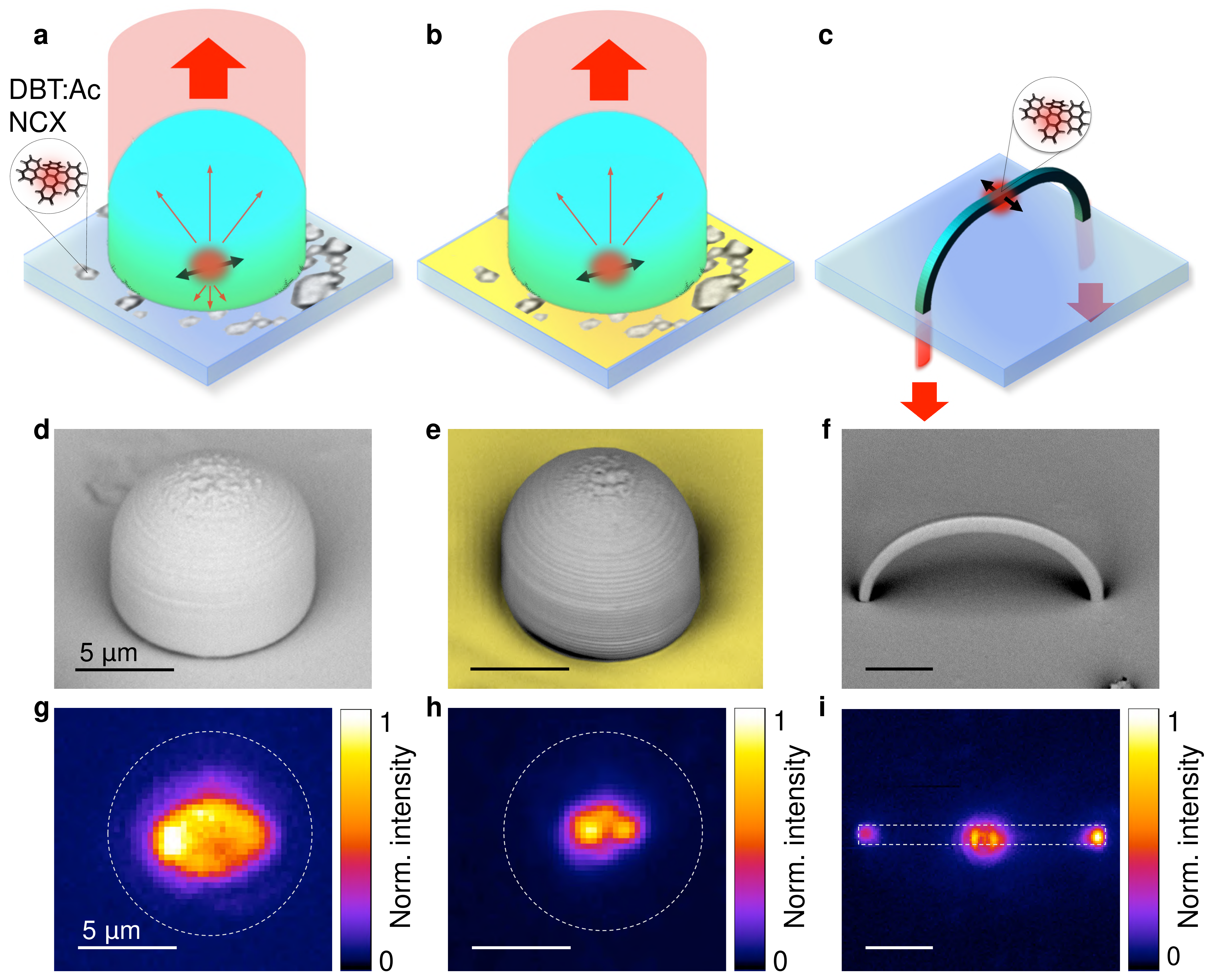}
\caption{\textbf{3D integration of organic quantum emitters in a polymeric platform}. \textbf{Top panel}, Artistic views of three light collecting devices realized with Direct Laser Writing and fully embedding an Anthracene nanocrystal with fluorescent molecules of Dibenzoterrylene (DBT:Ac NCX). The black double arrow represents the integrated emitter dipole and red arrows highlight the promoted direction for the emitted fluorescencce. \textbf{a},\textbf{b} Micro-dome lens design for upwards redirection, respectively realized on a silica substrate, and on a gold-coated layer to avoid bottom radiation and maximize collection efficiency. \textbf{c}, Suspended arch waveguide, fully enveloping the nanocrystal at its middlepoint, for efficient routing of fluorescence throughout the silica substrate. \textbf{Middle panel (d,e,f)}, Scanning Electron Microscope images of the respective structures. \textbf{Bottom panel (g,h,i)}, Fluorescence maps attesting the persistence of molecular emission after fabrication, acquired at room temperature and collected from the top in the micro-dome designs (\textbf{d},\textbf{e}) and through the substrate in the waveguide geometry (\textbf{f}). The dashed line in the maps denotes the contour of the respective structure.}
\label{fig:fig1}
\end{figure*}

Single PAHs are excellent candidates as non-classical light sources \cite{Lettow2010, Toninelli2010a, Rezai2019, Nicolet2007}, 
non-linear elements at the few photon level \cite{Hwang2009, Maser2016, Wang2019} and nano-scale sensors for electric fields, pressure and strain~\cite{TROIANI2019,Schadler2019,Tian2014,Mazzamuto2014}.
Recently they have been successfully integrated in open optical cavities \cite{Toninelli2010, Wang2019} and antennas \cite{Checcucci2017} and evanescently coupled to nanoguides\cite{Skoff2018, Faez2014} and waveguides\cite{Lombardi2017, Grandi2019, Turschmann2017}. 
Here, in Fig. \ref{fig:fig1} we present three different light collecting devices 3D-carved in polymers, each fully embedding a nanocrystal with fluorescent molecules. In the respective artistic views (top panels \ref{fig:fig1}a, \ref{fig:fig1}b, \ref{fig:fig1}c) the integrated emitter is represented by a dipole (black arrow) and red arrows highlight the promoted direction for the emitted fluorescence. Fig. \ref{fig:fig1}a shows a \textit{micro-dome} on a transparent dielectric substrate (silica, refractive index $n=1.45$). This is a modified Weierstrass solid immersion lens (SIL)~\cite{Barnes2002}, consisting of a hemisphere and a cylindrical base built on top of one DBT:Ac NCX. 
NCXs are previously deposited on the substrate (grey in the Figure and with a zoom on the DBT chemical structure) and selectively addressed upon preliminary characterization of the fluorescence intensity at room temperature. The structure is therefore centered on the emitter, so as to maximize upwards redirection and collection efficiency at narrow angles. In Fig. \ref{fig:fig1}b a similar design is sketched, this time realized on a gold-coated layer ($\sim200\,$-nm thick) to avoid radiation in the lower hemisphere. With DLW, suspended architectures can also be fabricated, such as the arch waveguide outlined in Fig. \ref{fig:fig1}c. In this case the nanocrystal is completely enveloped at the middlepoint of the structure, which is conceived to efficiently route fluorescence throughout the silica substrate. Besides the improvement in terms of integrated coupling efficiency~\cite{Shi2016}, the possibility of suspended integration is crucial to avoid losses through the substrate due to the usually low refractive index of polymer materials. The correct emitter position in the vertical dimension is ensured by the use of two photoresist layers with different viscosity but equal refractive indices. The accurate reproduction of the three designs can be appreciated in the scanning electron microscope images reported in Figs. \ref{fig:fig1}d, \ref{fig:fig1}e, \ref{fig:fig1}f. 

For each structure, fluorescence imaging has been acquired to evaluate the persistence of molecular emission. Figures \ref{fig:fig1}g, \ref{fig:fig1}h and \ref{fig:fig1}i represent the fluorescence maps obtained at room temperature, and collected from the top in the first two cases and through the substrate in the waveguide geometry (see Methods). 
The presence of fluorescence in all the devices in Fig. \ref{fig:fig1} is striking evidence for the compatibility of the organic molecular emitters with the whole fabrication process, both on dielectric and metallic surfaces and even in the suspended architecture. Notably, upon optimization of the writing parameters, all the investigated micro-domes showed molecule emission. The integration success rate is also enhanced by defining and anchoring the coordinate system of the writing laser beam to the location of selected nanocrystals within the optical microscope image. Deterministic positioning with such a simple and cost-effective fabrication method is key for scaling up the process to more complex chips, involving e.g. multiple emitters. 

We remark that the overall process of reprecipitation of the nanocrystals, deposition and selection of single photon sources, requires the experimenter less than one hour and can be readily automated.

The bright spots in Fig.\ref{fig:fig1}g and Fig.\ref{fig:fig1}h correspond to emission from the nanocrystals at the bases of the micro-domes. Comparing the two, a first indication is obtained of the higher directivity achievable in the presence of a reflective substrate underneath the SIL-like structure.  In Fig.\ref{fig:fig1}i instead, fluorescence is observed both in correspondence with the nanocrystal position at the centre of the waveguide, and at the lateral output ports, where the guided emission is outcoupled. From the relative intensities, a lower bound for the molecule-to-waveguide coupling efficiency can be hence estimated, yielding a value of about 10\%. We remark that this measurement is the result of the integrated signal from all the emitters in the NCX, each with different orientation and position with respect to the guided mode, hence with varying coupling efficiency. Furthermore, the additional propagation losses due to the emitter inclusion can be evaluated by comparing the normalized laser throughput ($\eta_{WG+NCX}=36\pm 1\,\%$) with the case of empty waveguides ($\eta_{WG}=40\pm 2\, \%\,$). The measured value for the additional scattering losses of $10\%$ can be further minimized upon selection of smaller nanocrystals, e.g. by integrating fluorescence imaging capability in the DLW workstation, as done e.g. in Ref. \cite{Shi2016}.

\begin{figure*}[ht!]
\centering
\includegraphics[width=0.85\textwidth]{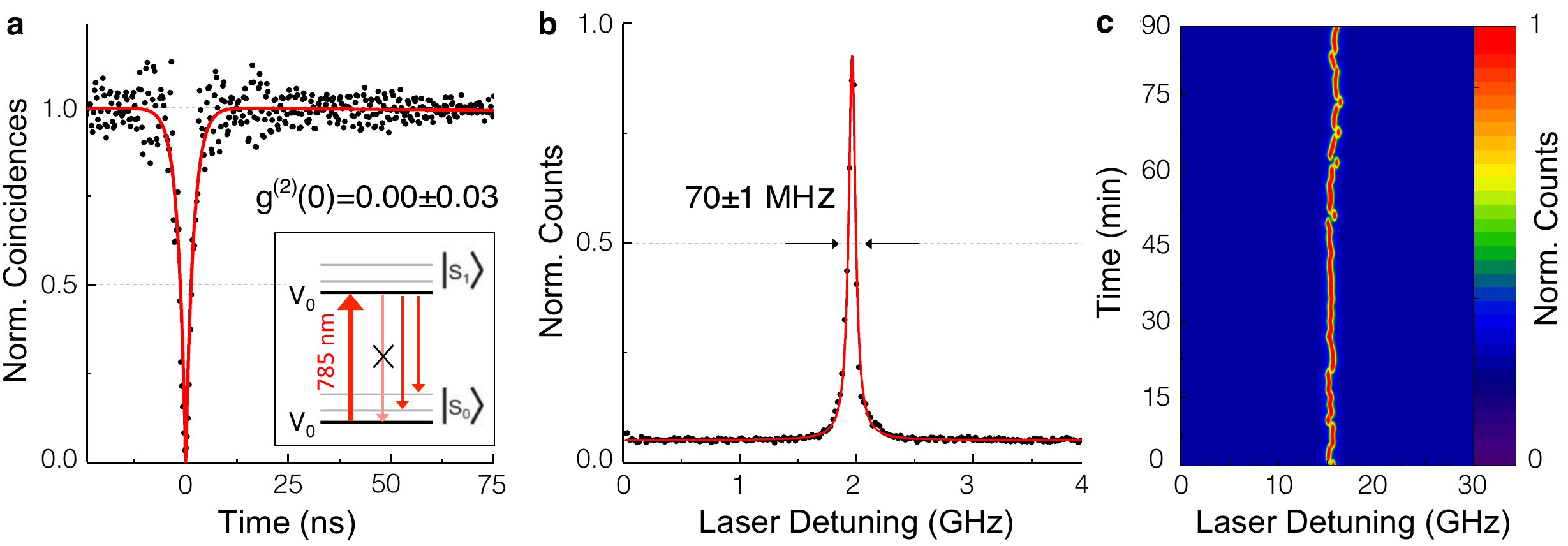}
\caption{\textbf{Optical characterization of the integrated emitters at cryogenic temperature}. Measurements are performed on the micro-dome on gold structure. \textbf{Inset}, Simplified Jablonski diagram of the pumping scheme. The zero-phonon line of a single DBT molecule is selected in frequency with a tunable continuous-wave laser ($\approx 785\,$nm) and the Stoke-shifted fluorescence is collected through a longpass filter. \textbf{a}, Histogram of photon coincidences on two avalanche photodiodes in a Hanbury Brown and Twiss configuration. The second order autocorrelation function is inferred by the best fit (red line) to the experimental data (black dots) $g^{(2)}(\tau)=1-(1-g^{2}(0))exp(-|\tau|\Gamma)$, $\tau$ being the time delay and $\Gamma$ accounting for the excitation and spontaneous emission rates, and yields $g^{(2)}(0)=0.00+0.03$. \textbf{b}, Excitation spectrum recording the Stoke-shifted fluorescence as a function of the laser frequency. The Lorentian fit (red line) to the experimental data (black dots) accounts for a near to the life-time limited line-width $\gamma_{exp}=70\pm1\,$ MHz. \textbf{c}, Photostability of the integrated molecule emission shown by repeated line-width measurements (\textbf{b}) over time for $1.5\,$h, with fluctuations of the excitation central frequency within the range of few line-widths.}
\label{fig:fig2}
\end{figure*}

In the following more accurate experiments are discussed, which characterize the emission at the single molecule level and in a cryogenic environment, using as test bed the geometry depicted in Fig. \ref{fig:fig1}b. Indeed, in the context of quantum technologies, coherence in the emission of single photons is required for basic quantum optics operations, whereas narrow lines enable sensing applications and efficient interfacing with atomic memories \cite{Siyushev2014}. Such properties can be observed with a standard epifluorescence microscope, upon cooling the sample down to $3\,$K (see Methods). The employed pumping scheme is shown in the simplified Jablonski diagram in the inset of Fig. \ref{fig:fig2}a. By scanning the frequency of a tunable continuous-wave diode laser around a wavelength of $\approx 785\,$nm, the zero-phonon line of a single DBT molecule within a highly doped nanocrystal can be readily selected. The Stoke-shifted fluorescence is collected through a longpass filter and analyzed in different aspects.\\
The emitted photon statistics is measured with two avalanche photodiodes (SPADs) arranged in a Hanbury Brown and Twiss (HBT) configuration. The second order autocorrelation function $g^{(2)}(\tau)$ is inferred from the histogram for short delay times of the photon coincidences on the two SPADs. A typical result from the micro-dome on gold structure is displayed in  Fig. \ref{fig:fig2}a, obtained at saturation. The best fit to the experimental data with the function $g^{(2)}(\tau)=1-(1-g^{(2)}(0))exp(-|\tau|\Gamma)$, where $\tau$ is the time delay and $\Gamma$ accounts for the excitation and spontaneous emission rates, yields $g^{(2)}(0)=0.00+0.03$. Such value clearly verifies single molecule emission and proves the high purity of the single photon stream emitted in the integrated-molecule geometry.
\\
By recording the Stoke-shifted fluorescence as a function of the laser frequency, operating well below saturation, the excited state population is probed and the zero-phonon line-width can be directly retrieved. A characteristic excitation spectrum is shown in Fig. \ref{fig:fig2}b with a near to the life-time limited line-width $\gamma_{exp}=70\pm1\,$MHz. Considering that $\gamma_{exp}$ is within the experimental statistics collected for the non integrated DBT:Ac NCX~\cite{Pazzagli2018}, and that it is close to the theoretical Fourier-limited value of $40\,$MHz for DBT molecules, we can deduce that this fabrication process does not alter significantly the NCX properties and that quantum coherence is not compromised. \\
Finally, spectral diffusion is investigated by repeating the line-width measurement previously described over time, for $1.5\,$h. Fig. \ref{fig:fig2}c shows a good photostability of the integrated molecule emission, whose excitation central frequency fluctuates within the range of few line-widths. Being spectral diffusion a sensitive tool to appreciate the environmental disturbance on quantum emitters, these results are a relevant indication that the proposed polymeric platform is a reliable technique to embed single molecules in a photonic circuit.\\

\section*{Record-high photon flux from a single molecule at low temperatures} \label{CaseStudy}
We choose the micro-dome design as a case study to investigate its performances in terms of light extraction from single molecules. The total height of the structure is given by $h=(1+1/n)r$, where $n$ is the polymer refractive index and $r$ the radius of curvature. This design provides redirection via refraction at all upwards angles and efficiently compresses emission into a small numerical aperture. 
It also offers advantages in terms of simplicity of fabrication and robustness against imperfections. 

\begin{figure*}[ht!]
\centering
\includegraphics[width=0.76\textwidth]{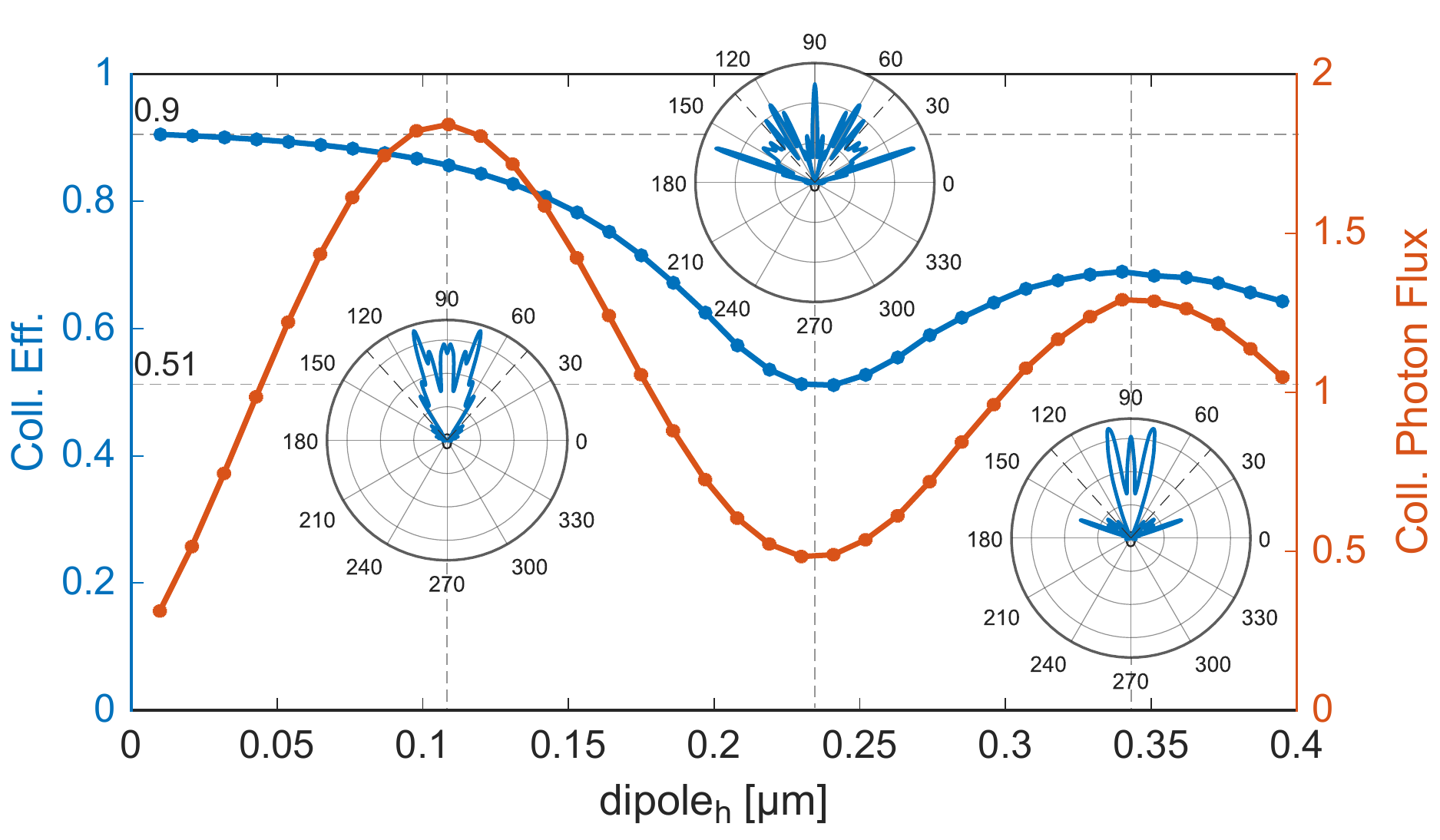}
\caption{\textbf{2D numerical simulation of the micro-dome geometry}. Collection efficiency and collected photon flux calculated as a function of the fluorophore vertical position within the nanocrystal volume (dipole$_h$), for the the micro-dome on gold. The collected photon flux is calculated as the radiated power within the numerical aperture acceptance $NA\sim0.62$ ($P_{hom}$) and the total power emitted in a homogeneous medium ($P_{hom}$). For dipole$_h=110\,$nm the collection efficiency is maximum, above $85\%$, and the corresponding polar plot shows a highly directional pattern with $\sim80\%$ of the total emission within a polar angle of $\sim30^{\circ}$. While the emission is quenched close to the metal layer, at bigger distances the overall collection efficiency is reduced due to emission lobes beyond the collection limits (grey dashed lines in the polar plots), as displayed in the corresponding emission patterns.}
\label{fig:fig3}
\end{figure*}

In Fig. \ref{fig:fig3}, the results of 2D numerical simulations for the collection efficiency and the collected photon flux (for the experimental numerical aperture $NA\sim0.62\,$) for the micro-dome geometry on gold are summarized (see Methods). The structure dimensions are optimized assuming a dipole-to-gold distance dipole$_h=100\,$nm, which promotes emission in the upper hemisphere following Ref. \cite{Checcucci2017}, yielding $h_{th}=8.7\,\mu$m and $r_{th}=5.3,\mu$m. 
Since experimentally though there is no control over the fluorophore position within the nanocrystal volume, calculations are performed varying its value within the largest possible NCX thickness of $400\,$nm. The DBT alignment within Ac crystal is instead well known and lies horizontally on the plane of the substrate\cite{Toninelli2010a}. 
As a result, for dipole$_h\approx100\,$nm, a remarkable collection efficiency above $85\%$ is obtained, together with a small enhancement of the total emission rate. In the corresponding polar plot, the emission pattern appears highly directional, resulting in $\sim 80\%$ of the total emission funneled within a polar angle of about $30^{\circ}$.
The oscillatory behavior in the detected photon flux is due to the interference nature of the photonic redirection effect, as already discussed for a similar case e.g. in Ref.\cite{Checcucci2017}. Close to the metal layer, the dipolar emission is obviously quenched. As the dipole position is further displaced away from the gold surface, the onset of emission lobes beyond the critical angles is clearly visible in the second polar plot, reducing the overall collection efficiency, even when constructive interference in the forward direction is recovered for dipole$_h\simeq 350\,$nm.

\begin{figure*}[ht!]
\centering
\includegraphics[width=0.85\textwidth]{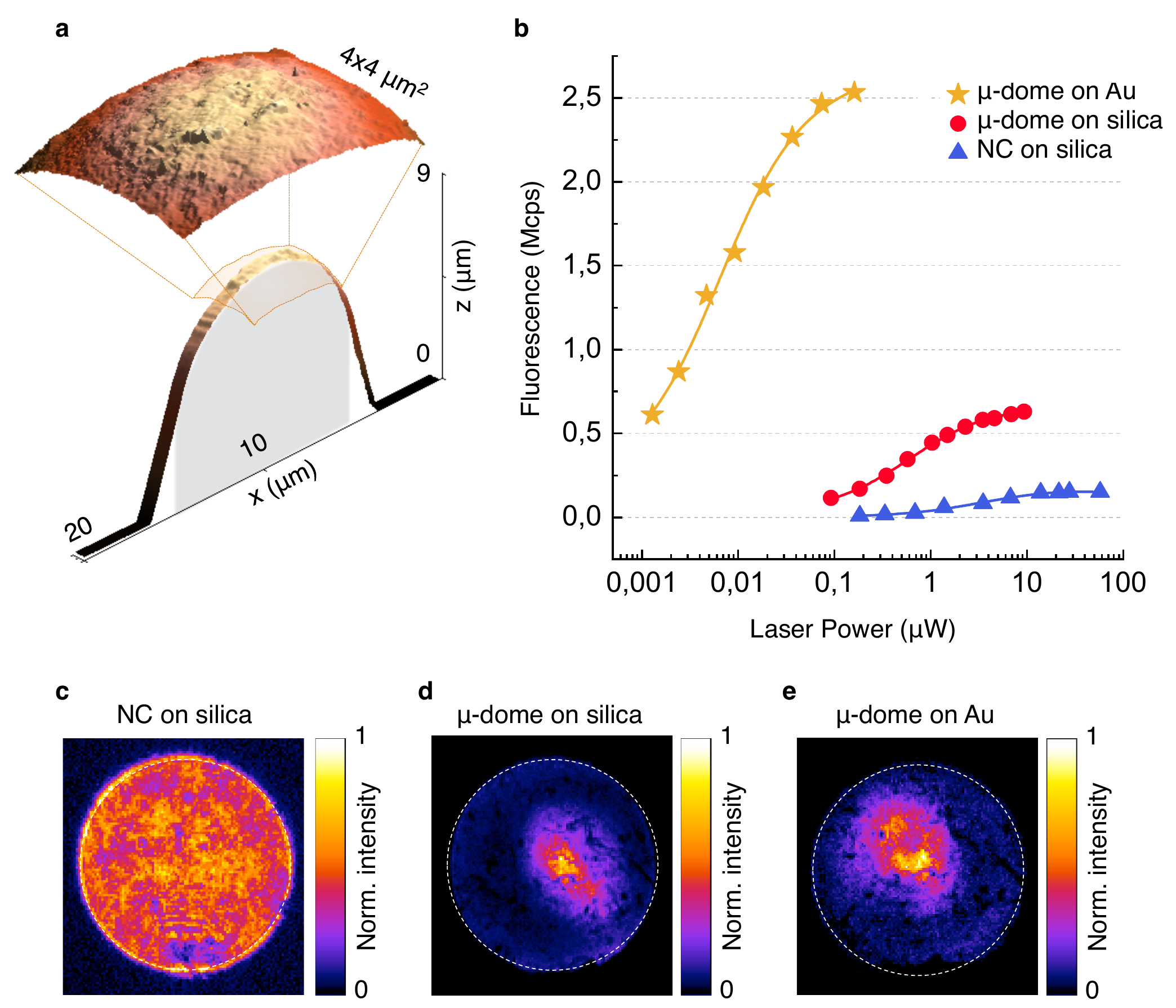}
\caption{\textbf{Experimental results}. \textbf{a}, Atomic Force Microscopy (AFM) characterization of the fabricated micro-dome on gold. The strip-scan ($20$x$1\,\mu$m$^2$) of the structure passing through the dome pole gives an estimation of the experimental dimensions, with a total height $h=9.1\pm0.1\,\mu$m and a radius $r=5.0\pm0.1\,\mu$m, in good agreement with the optimal values based on simulations, $h_{th}=8.7\,\mu$m and $r_{th}=5.3\,\mu$m, and displayed in the grey silhouette. From the AFM map of the top $4$x$4\,\mu$m$^2$ surface area, the surface roughness is estimated to be $\Delta r = 200\pm150\,$nm. \textbf{b}, Experimental collected photon flux at low temperature from a single molecule embedded in the fabricated micro-dome structure on gold (yellow stars), compared to the case of the micro-dome on silica (red dots) and of the nude nanocrystal (blue triangles). Saturation curves are obtained by plotting the maximum counts of the measured excitation spectrum as a function of pump power. The data are fitted (continuous lines) according to the formula $R(P)=(a+bP)+R_{\infty}P/(P+P_s)\,$. The corresponding maximum detected photon flux at the detector are $R_{\infty_{Au-dome}}=(2.38\pm0.11)\,$Mcps, $R_{\infty_{ dome}}=(0.66\pm0.02)\,$Mcps and $R_{\infty_{NC}}=(0.18\pm0.01)\,$, respectively. \textbf{c,d,e}, Back focal plane (BFP) measurements of the emitted fluorescence showing the emission pattern for the micro-dome on gold, the micro-dome on silica and the nude nanocrystal, respectively. Detected counts are plotted in momentum space and normalized to the maximum for all plots, whereas the white dashed circumferences identify the experimental maximum collection angle of $\sim 42^{\circ}\,$.}    
\label{fig:fig4}
\end{figure*}

Based on such theoretical results, the fabrication of the micro-dome on gold embedding a single DBT:Ac NCX was optimized and characterized by means of atomic force microscopy (AFM). A typical AFM image is displayed in Fig. \ref{fig:fig4}a. In particular, a strip-scan of the structure passing through the dome pole is shown together with a high resolved AFM map of the top $4$x$4\,\mu$m$^2$ surface area. From the dome profile we estimate the experimental dimensions of the structure, with a total height $h=9.1\pm0.1\,\mu$m and a radius $r=5.0\pm0.1\,\mu$m, in good agreement with the nominal values set by the numerical optimization. From the AFM map, the average surface roughness is estimated to be $\Delta r=41\pm4\,$nm. 

Finally, Fig. \ref{fig:fig4}b shows one of the main results of the paper, that is the collected photon flux as a function of pump power at low temperatures, emitted from a single molecule embedded in the polymeric micro-dome device, whose design has just been commented. The measurements are compared to the case of a nude nanocrystal and of a molecule in the micro-dome device on silica. The photon flux is obtained, for a given excitation power, by measuring the excitation spectrum and recording the detected count rates in the Stokes-shifted band at resonance, without correcting for the detector internal efficiency, ($\eta_{det}\simeq65\,\%$). In case indistinguishable photons from the ZPL are required, non-resonant excitation on low-density NCXs can be employed \cite{Lombardi2019} and a similar count rate can be estimated, based on the typical expected branching ratio for DBT:Ac ($30\% \div 50\%$ \cite{Wang2019}). For each configuration, the variability in the emitter position within the crystal results in some statistical fluctuations in the measured count rate within nominally identical structures. The stronger effect is to be found for the dome-on-gold, as follows from the previous discussion on Fig. \ref{fig:fig3}. The direct comparison reported in Fig. \ref{fig:fig4} is however representative of the typical observed trend. 

As expected, a saturation curve is observed in all three cases as a function of the pump power, expressed in logarithmic scale. The detected counts are fitted according to the formula $R(P)=(a+bP)+R_{\infty} P/(P+P_s)\,$, with $R_{\infty}$ being the maximum detected count rate, $P$ the laser power, $P_s$ the saturation power and $(a+bP)$ accounting for the residual laser scattering as a linear contribution. We obtain a maximum measured photon flux at the detector equal to $R_{\infty_{NC}}=(0.18\pm0.01)\,$Mcps, $R_{\infty_{ dome}}=(0.66\pm0.02)\,$Mcps and $R_{\infty_{Au-dome}}=(2.38\pm0.11)\,$Mcps, respectively. The detected count rate from a single cold molecule in the optimized micro-dome design is to our knowledge the highest reported to date and appears particularly significant when associated with the long excited state lifetime of DBT molecules ($\simeq 4.5$ $ns$). Furthermore, taking into account the efficiency of the detector ($\eta_{det}$), of the optical path ($55\%$) and the objective transmission ($70\%$), the real photon flux at the detector is $\sim10\,$Mcps. In particular,
%
with respect to the nude nanocrystal, we measure a factor $3.7\pm0.2$ improvement in the collected photon flux for the micro-dome on silica, and a factor $13\pm1$ in the case of the optimal configuration for the micro-dome on gold. Moreover a reduction in the saturation power by about two orders of magnitude is observed, passing from a value of $(56\pm10)\,$W/cm$^2$ to $(0.10\pm0.02)\,$W/cm$^2$. Indeed, the presence of the modified SIL has the primary effect of increasing the effective numerical aperture of the optical system, hence improving its focusing efficiency. The reflective layer underneath further increases the intensity at the emitter position for a given input power. 

In order to probe the emission pattern of the devices, we perform back focal plane (BFP) measurements of the emitted fluorescence, providing the power distribution in momentum space. In Figures \ref{fig:fig4}c, \ref{fig:fig4}d, \ref{fig:fig4}e we compare the normalized BFP maps for the nude nanocrystal on silica, the micro-dome device on silica, and the micro-dome on gold. The white dashed circumferences in the plots identify the experimental maximum collection angle of $\sim42^{\circ}$, whereas the centre of the circumference corresponds to the objective optical axis. Fig. \ref{fig:fig4}c shows a homogeneously bright map, as expected from a dipole emission and consistently with the typical on-plane orientation of the DBT:Ac system~\cite{Toninelli2010a, Pazzagli2018}. Conversely, Fig. \ref{fig:fig4}d and \ref{fig:fig4}e confirm an effective redirection into small angles achieved in the micro-devices, competitive with what shown in much more complex structures~\cite{Johlin2018}. In the last cases, the slight misalignment of the bright spots is evidence of the not perfectly centred emitter, which owes both to the 3DLW process and to the uncertainty in the fluorophore position within the Ac nanocrystal. Consistent with the results of the numerical simulations presented in Fig.\ref{fig:fig3}, the pattern of a micro-dome on gold hosting a horizontally oriented molecule at an appropriate distance from the gold surface ($\simeq 100 nm$) features most of its emission within the first $30^{\circ}$ (see also Supplementary Information (S.I.) Fig.1). It is worth noticing that this configuration is not the one exhibiting the highest directivity. Narrower emission lobes can indeed be observed (see Fig.2 in the S.I.), corresponding though to a lower collection efficiency, as clearly evident from the central polar plot of Fig.\ref{fig:fig3}. 



\begin{figure*}[ht!]
\centering
\includegraphics[width=0.85\textwidth]{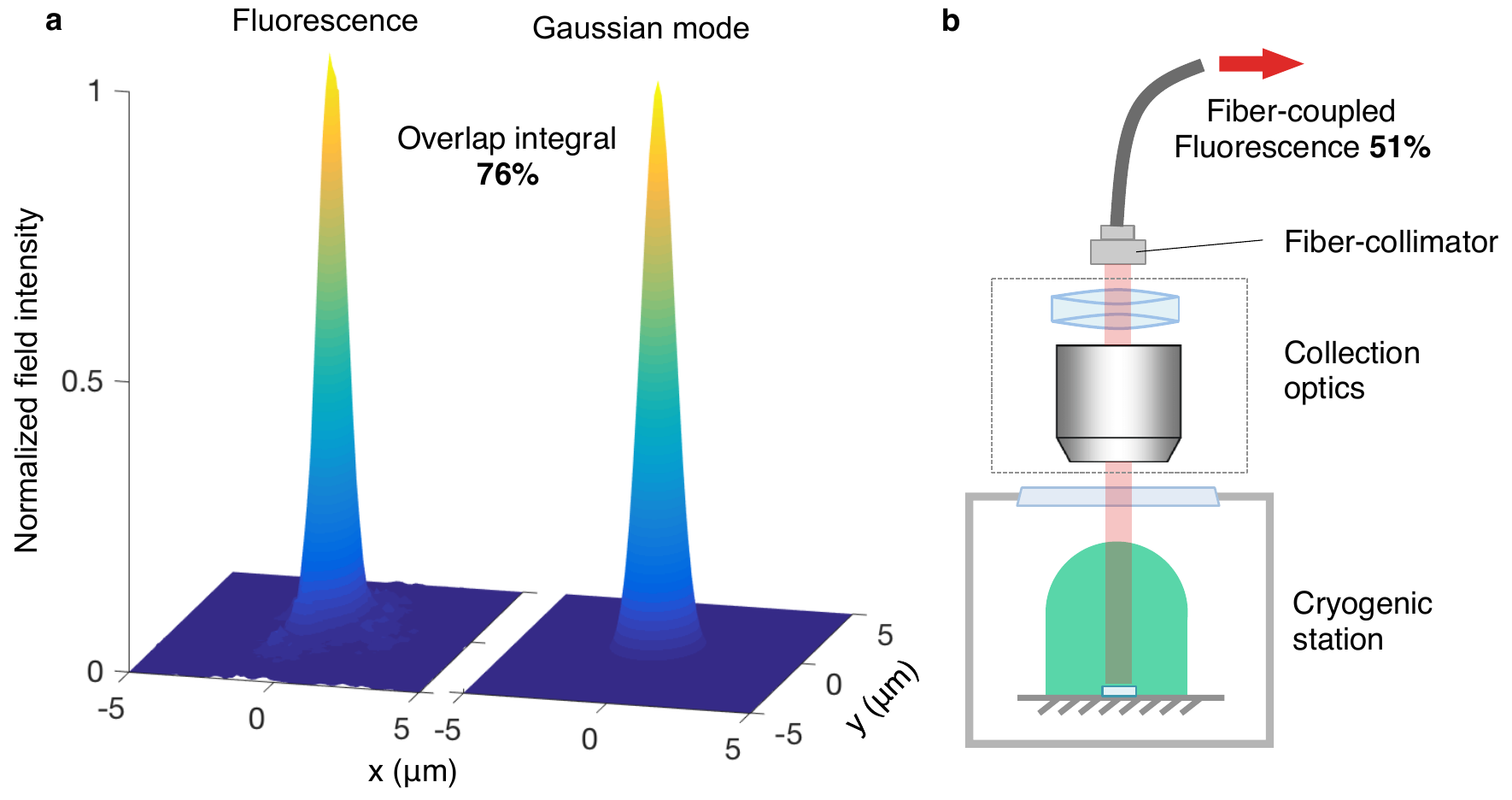}
\caption{\textbf{Efficient coupling into a fiber}. \textbf{a}, Estimation of the expected coupling efficiency for the optimal configuration of the micro-dome on gold. The target value is evaluated in terms of mode matching by comparison of the measured fluorescence spatial distribution of the integrated structure with a Gaussian mode, typical of a single mode optical fiber. The estimation gives the best overlap integral of $\approx 80\%\,$. \textbf{b}, Schematic of the experimental setup and measured performance of up to $51\%$ coupling efficiency, which is normalized to the detected counts on a SPAD. In particular, the setup consists in the objective, a telescope and a tunable-distance fiber-collimator. }
\label{fig:fig5}
\end{figure*}
The effective redirection at narrow angles combined with the high detected count rate achieved with the micro-dome device are particularly suitable for the efficient coupling into a fiber, within the perspective of a compact fiber-coupled single photon source. 
With the experimental setup schematically shown in Fig. \ref{fig:fig5}b, a coupling efficiency up to $51\%\,$ into a single mode fiber is achieved.
The target value can be estimated in terms of mode matching by comparing the measured fluorescence spatial distribution in the micro-dome on gold with a Gaussian mode, typical of a single mode optical fiber (see Fig. \ref{fig:fig5}a). 
In particular, we estimate the best overlap integral to be of the order of $\simeq80\,\%$. Considering the flexibility offered by a polymeric photonic platform \cite{Gissibl2016}, a monolitic fiber coupler can be integrated onto the metal-coated micro-dome in order to realize an alignment-free and ultra compact fiber-coupled SPS. 
\section*{Conclusions}
We have demonstrated the fast, efficient integration of single molecule quantum emitters in 3D polymeric structures, thus realizing a novel flexible photonic platform, leveraging the versatility of polymeric materials. The experimental results entail deterministic positioning of the source, fabrication on different substrates (dielectric, metallic) as well as integration in suspended designs. Furthermore, the compatibility of direct laser writing with organic nanocrystals is demonstrated by showing close-to Fourier-limited emission from a single embedded dibenzoterrylene molecule at $3\,$K.
As a first example of the performances attainable by our platform, we show unequalled collection efficiency for single organic emitters at low temperatures, reaching photon fluxes above $2.3\,$ Mcps. 
This value approximately doubles the state of the art for the detected count rate from a single molecule, with the improvement rising to a factor 5 considering the specific case of DBT \cite{Polisseni2016, Wrigge2008}. The effective fluorescence redirection displayed in back-focal-plane measurements makes this device suitable for fiber-coupling, with an expected efficiency of $\simeq80\%$ and a measured value up to $50\%$. 

The use of single-molecule emitters, intrinsically identical, offers the appealing possibility of conveying indistinguishable photons from different remote sources \cite{Lettow2010}.
Few-photon non-linearities can be also envisaged, based on such quantum emitters interacting with confined optical modes \cite{Kewes2016, Wang2019}.
Owing to the great flexibility of the direct laser writing technique and to the versatility of polymeric materials, the proposed approach can be potentially extended to operational chips integrating monolithic cavities, waveguides and directional couplers~\cite{Flatae2015}. For example, a complex quantum effect such as the interference among more than two particles can be observed, using on-demand single photons generated by individual molecules embedded in 3DL-written polymeric tritters \cite{Spagnolo2013}. Otherwise, the proposed technology can be adapted for scalable and efficient coupling of multiple emitters, even in hybrid polymer-silicon nitride circuits~\cite{Schumann2014}. 
Furthermore, the possibility to engineer photo-functional materials as photoresists ~\cite{Kudo2009} 
enables full customization, from the optimization of spatial resolution and mechanical stability, to the employment of photo-reactive ~\cite{nocentini2018} or electro-optical features for tunable resonances. Thanks to its advantages, we expect that our platform will lead to a plethora of novel scalable micro-photonic devices with tailored functionalities, at the single emitter level. Our achievements anticipate the impact organic materials can have also in quantum technologies.

\subsection*{Acknowledgements}
\vspace{-0.2cm}
The authors acknowledge and thank Hengsbach Stefan and Bade Klaus from the Institute of Microstructure Technology (MIT) in Karlsruhe, for the useful discussions and insights on the fabrication process. 

\subsection*{Funding}
\vspace{-0.2cm}
This project has received funding from the EraNET Cofund Initiatives QuantERA
within the European Union's Horizon 2020 research and innovation program grant agreement No. 731473 (project ORQUID). M.C. acknowledges the Karlsruhe Nano Micro Facility (KNMF) for the access to fabrication and characterization technologies within the KNMF proposals 2018-019-021009 and 2019-021-025773. 

\subsection*{Author contributions}
\vspace{-0.2cm}
M.C., P.L. and C. T. conceived the research and designed the experiments. M.C. fabricated the devices with help from S.N.. M.C, P.L. and M. T. carried out the experimental characterization. F.P., P.L., M.C. carried out the theoretical analysis. B.T. and M.C. performed the morphological characterization. C.T. supervised the project. All authors discussed the results. M.C. and C.T. wrote the manuscript with critical feedback from all authors.

\subsection*{Competing financial interests}
The authors declare no competing financial interests.





%
%
%

\bibliographystyle{naturemag}
\bibliography{ref}



\section*{Methods}

\subsection*{Sample Fabrication}
The Direct laser writing process is based on two-photon polymerization induced by a strongly focused laser beam. We employed a commercial DLW workstation (\textit{Nanoscribe}) and the commercially available negative-tone IP-photoresists (IPs). IPs provide high resolution and high mechanical stability for structures in the micro- and sub-micron range and are chemically compatible with the Ac crystalline matrix. Moreover they require low exposure and developing time for the writing process~\cite{Mueller2014,Jiang2014}, thus minimizing possible detrimental effects on DBT:Ac NCXs. Finally, the choice of IPs is motivated by the absence of background fluorescence in the spectral range of the molecule emission.\\
The DBT:Ac NCX growth procedure consists in injecting $250\, \mu$L of a  mixture of $4:10^3$ $1\,$mM DBT-toluene and $5\,$mM Ac-acetone solutions into $5\,$mL of sonicating water. While continuously sonicating the system for $30\,$min, solvents dissolve in water and DBT:Ac crystals are formed in aqueous suspension, as described in detail in ref.~\cite{Pazzagli2018}. Solvents and Ac were purchased from Sigma-Aldrich, water was deionized by a Milli-Q Advantage A10 system (resistivity of $18.2\,$m$\Omega\, \cdot\,$cm at $25\,^{\circ}$C), and DBT was purchased from Mercachem.\\
In the micro-dome configurations, the nanocrystals were deposited on a fused silica coverslip via dessication after drop-casting $\sim 10\, \mu$L of the aqueous suspension. Before the deposition, the substrate was extensively cleaned with acetone and optical paper. In the micro-dome on gold configuration, the substrate was also sputter-coated with a $\sim200\,$nm thick gold film. After the deposition NCX were protected from sublimation and from fabrication damage by spin-coating a $~200\,$nm thick layer of PVA (purchased from Sigma-Aldrich). To fabricate the micro-lens structures, the optimized geometry is translated into a series of concentric rings by an automated script which fits the volumetric dimensions of the writing laser focus (the voxel, typically ellipsoidal and depending on laser power) to the target external surface dimensions and smoothness of the structure. We used $7.5\,$mW laser power, measured at the entrance of the objective. The translated rings were written directly in drop-cast IP-DIP (micro resist technology GmbH, $n=1.54$) resist via dip-in DLW lithography (Nanoscribe GmbH). Each micro-dome was written on the top of a single nanocrystal by localizing it through the same microscope objective at the imaging feedback of the workstation. Total writing time for one micro-lens is $\sim20\,$min. The unexposed photoresist was eventually removed in a development bath of propylene glycol monomethyl ether acetate (PGMEA) for $5\,$min and then in isopropyl alcohol for $5\,$min. After development the sample was carefully dried with a weak clean air flux.\\
For the suspended waveguide configuration, DBT:Ac NCX were dispersed in the photoresist such to be successively embedded in the suspended middlepoint of the structure. This is achieved by firstly spin-coating a layer of IP-G (micro resist technology GmbH, $n_=1.52$) resist of $5\div10\,\mu$m thickness, upon which NCX are then deposited. This thickness is imposed by mechanical stability (waveguides bend and fall for thickness $\gtrsim10\,\mu$m) and by coupling efficiency (fluorescence is absorbed by the substrate and not efficiently guided for thickness $\lesssim5\,\mu$m) reasons. The choice of IP-G is prescribed by the necessity of rigidity of the spin-coated layer in order to avoid sinking of the deposited NCX, and this is attained by pre-baking of IP-G at $100\,^{\circ}$C for $30\,$min. Therefore, deposition of NCX was carried through the same procedure previously described, and followed by drop-casting of IP-DIP for total envelopment of the emitters. The suspended waveguide was then written in one step fabrication by localizing the nanocrystal, estimating its distance from the substrate via the piezo-system feedback of the workstation, and defining the suitable waveguide dimensions to be employed to effectively wrap the nanocrystal. In this case, the waveguide design was translated into a series of arch-like lines. Development and drying was performed as previously described.

\subsection*{Morphological characterization}
Topological images were performed by SEM (Phenom Pro, PhenomWorld) and AFM (home-built equipment based on piezo XYZ positioning system (PI-527.3CL from Physik
Instrumente) and R9 controller (RHK Technology)) equipped with a soft silicon probe CSG01 (NTMDT). In particular, for SEM contrast images, after optical characterization, the samples were sputter-coated with a $\sim50\,$nm gold film.
\subsection*{Simulations}
Numerical simulations of the collection efficiency for non-integrated nanocrystal on silica and for the NCX inside the micro-domes on silica and on gold respectively, were performed using a commercial software (Comsol Multiphysics) implementing the 2D Finite Element Method (FEM). The nanocrystal is modelled in each configuration as a $800\,$nm$\,$x$\,400\,$nm platelet of Ac ($n=1.7$) with the molecules modelled with an in-plane dipole source at the centre of the nanocrystal, $100\,$nm from the substrate. A layer of $200\,$nm of PVA ($n=1.475$) covers the whole considered area, including the nanocrystal.
In the integrated configurations the micro-dome consists in a hemisphere of radius $5.3\,\mu$m placed on a cylindrical base with the same radius and height $h=3.4\, \mu$m. 
For the optimal configuration on reflective surface, the silica substrate is covered with a $180\,$nm-thick layer of gold. 
The collection efficiency is evaluated by integrating the far fields within the acceptance angle defined by the numerical aperture of our setup ($42^{\circ}$), and normalizing by the integrated far fields on the whole $2\pi$ angle. 
The collected photon flux is calculated as the ratio between the power radiated by a dipole in the far field within the maximum acceptance angle given by the numerical aperture $NA\sim0.62\,$ and the power radiated by the same dipole in a homogeneous medium with the refractive index of $1.7$.
\subsection*{Optical Set-up}
The optical characterization of DBT molecules within the Ac NCX was performed with a versatile home-built scanning fluorescence confocal microscope. The set-up is equipped with a closed-cycle helium cryostat (Cryostation by Montana Instruments), capable of cooling samples to $2.7\,$K. Molecules can be excited off-resonance at $767\,$nm by a CW laser (Toptica DL110-DFB) or alternatively at cryogenic temperature with a resonant tunable narrowband CW laser (Toptica, LD-0785-0080-DFB-1), which is centered at $784.6\,$nm and can be scanned continuously in frequency over a range of $800\,$GHz. All laser sources are fibercoupled and linearly polarized by means of a half-wave plate in the excitation path to allow optimal coupling to single DBT transition. The laser intensities reported in the main text are calculated from the power measured at the objective entrance divided by the measured area of the confocal spot at the substrate. For low-temperature measurements, the excitation light is focused onto the sample by a glass-thickness- compensation air objective (OptoSigma $50\,$x, $N.A.=0.65$, $WD=10.48\,$mm) and can be scanned over the sample by a telecentric system and a dual-axis galvo-mirror. For room temperature measurements, a long-distance air objective (Mitutoyo $100\,$x Plan Apochromat, $N.A. = 0.7\,$, $WD = 6\,$mm) is used to focus light onto the sample, which is mounted on a piezoelectric nanopositioner (NanoCube by Physik Instrumente). The Stokes-shifted fluorescence is collected by the same microscope objective used in excitation, separated from the excitation light through a dichroic mirror (Semrock FF776-Di01) and a long-pass filter (Semrock RazorEdge LP02-785RE-25), and detected by either an EMCCD camera (Andor iXon 885, $1004\,$x$1002\,$pixels, pixel size $8\,\mu$m$\,$x$\,8\, \mu$m) or by a SPAD ($\tau$-SPAD-50 single photon counting modules by PicoQuant). For the HBT configuration, we employ two fibred SPADs (fiber-coupled photon counting modules by Excelitas) and a time correlated single photon counting (TCSPC) card (PicoHarp, PicoQuant). A converging lens can be inserted in the excitation path to switch between confocal and wide-field illumination, while a converging lens is added in the detection path before the EMCCD camera to study the wave-vector distribution of the light emitted by single DBT molecules via BFP imaging. For the fiber-coupling configuration, a single-mode fibre, a fiber-collimator and a telescope were employed to match the fluorescence mode to the fibre. Measurements of laser throughput in the waveguides were performed by focusing the laser at one coupler, through the substrate, and evaluating the transmitted light at the opposite coupler. The measured signal was then normalized to the direct reflection of the input laser on a mirror. Finally, the value was corrected for the overlap integral between the laser spatial mode and the waveguide spatial mode, which were not optimally matched in the experiment.

\end{document}